# Equatorial jet in the lower to middle cloud layer of Venus revealed by Akatsuki

This document contains the original manuscript (submitted version) of a paper published online from *Nature Geoscience* on 28 Augst 2017:

T. Horinouchi, S. Murakami, T. Satoh, J. Peralta, K. Ogohara, T. Kouyama, T. Imamura, H. Kashimura, S. S. Limaye, K. McGouldrick, M. Nakamura, T. M. Sato, K. Sugiyama, M. Takagi, S. Watanabe, M. Yamada, A. Yamazaki & E. F. Young (2017) Equatorial jet in the lower to middle cloud layer of Venus revealed by Akatsuki. *Nature Geoscience*, Vol. 10, doi:10.1038/ngeo3016.
http://www.nature.com/ngeo/journal/vaop/ncurrent/full/ngeo3016.html

The main differences between the manuscript and the final published paper are as follows:

- The vertical region mainly sensed by the nightside observation by the IR2 instrument (alt: 45-60 km) is called as the lower and the middle cloud regions in the paper, but it is called the lower cloud layer in the manuscript. Accordingly, the title was slightly modified.
- In the manuscript, the vertical region is simply assumed, but in the paper, a substantial discussion is made to show that the jet feature is likely an horizontal (meridional) structure. It includes an addition of a figure.
- An author was added for the contribution to the above discussion (Kevin McGouldrick)

# Equatorial jet in the lower cloud layer of Venus revealed by Akatsuki

Takeshi Horinouchi[1]*, Shin-ya Murakami[2], Takehiko Satoh[2,3], Javier Peralta[2], Kazunori Ogohara[4], Toru Kouyama[5], Takeshi Imamura[6], Hiroki Kashimura[7], Sanjay S. Limaye[8], Masato Nakamura[2], Takao M. Sato[2], Ko-ichiro Sugiyama[9], Masahiro Takagi[10], Shigeto Watanabe[11], Manabu Yamada[12], Atsushi Yamazaki[2], Eliot F. Young[13]

Venus is covered with thick clouds that extend from ~45 to ~70 km altitude above the mean surface, but atmospheric thermal radiation is emitted to space at narrow spectral windows of near infrared. As the near-infrared radiation is partially attenuated by inhomogeneous clouds in the lower cloud layer, it is possible to measure winds by tracking cloud silhouettes seen on the night-side. The Venusian atmosphere is known for the super-rotating westward wind much faster than planetary rotation. Past studies reported that the westward wind speed in the lower levels of the cloud layer ranges 50 to 70 m/s at low-to-mid latitude,peaking at mid-latitude or having a nearly constant meridional profile. Here we report the detection of a jet-like wind maximum at low-latitude with unprecedented high speed greater than 80 m/s, with a peak angular speed around the planetary rotation axis peaking near the equator, a feature that has not been found previously at any altitude. The finding is made by tracking features seen in the IR2 camera images from Akatsuki orbiter taken during July–August 2016. The mechanism of the equatorial is yet to be elucidated, and it is expected to provoke further studies of the superrotation of the Venusian atmosphere.

[1] Faculty of Environmental Earth Science, Hokkaido University, N10W5, Sapporo, Hokkaido 060-0810, Japan
[2] Institute of Space and Astronautical Science, Japan Aerospace Exploration Agency
[3] Department of Space and Astronautical Science, School of Physical Sciences, SOKENDAI
[4] School of Engineering, University of Shiga Prefecture
[5] Artificial Intelligence Research Center, National Institute of Advanced Industrial Science and Technology
[6] Graduate School of Frontier Sciences, the University of Tokyo
[7] Department of Planetology / Center for Planetary Science, Kobe University
[8] Space Science and Engineering Center, the University of Wisconsin-Madison
[9] National Institute of Technology, Matsue College
[10] Division of Science, Kyoto Sangyo University
[11] Space Information Center, Hokkaido Information University
[12] Planetary Exploration Research Center, Chiba Institute of Technology
[13] Southwest Research Institute

The planet Venus rotates westward with a very low angular speed corresponding to a period of 243 days, but its atmosphere rotates to the same direction with much higher angular speeds[1]. This superrotation reaches its maximum near the cloud top located at around the altitude of 70 km, where the rotational periods are 3 to 5 days, several tens of times faster than the planetary rotation[1,2]. Measurements by entry probes like Veneras, Pioneer Venus Multiprobe, and VEGA revealed that zonal wind speeds below the cloud top decreases quasi-linearly with depth[2--4]. Despite the long history of studies[5--10], the mechanism of the superrotation remains unsolved.

The near-infrared window between the wavelengths of 1.7 and 2.4 μm and its applicability for the lower-cloud tracking was first explored with ground-based observations[11], and later during the Galileo flyby[12] and with the Venus Express orbiter[13,14]. These and other previous studies consistently set the zonal wind speeds in the lower cloud layer in the range of 50 to 70 m/s over low to mid latitude, weakly peaking at mid-latitude[11--17] . In contrast, Crisp et al[18] performed a two-month campaign of ground-based observations in 1990 matching Galileo's Venus flyby, finding that features with horizontal scales greater than 2,000 km tend to move faster than smaller-scale features. These large-scale features moved roughly at a uniform angular velocity (solid-body rotation) and westward speeds around 80 m/s at the equator. They speculated that the movement of the large-scale features reflects the winds in the middle cloud layer at around 55 km, while the small-scale features are advected by the winds in the lower cloud layer. The speculation, however, has not been confirmed by later studies.

Akatsuki was launched in 2010, and after the failure of the first orbit insertion in the same year, it was successfully maneuvered to start orbiting Venus on Dec 7, 2015[19,20]. Its orbital period is about 11 days, and its low orbital inclination (<10°) makes it suitable to observe low latitudes. Akatsuki is equipped with imaging instruments designed to observe the Venusian atmosphere and clouds at multiple wavelengths. One of them, the Longwave Infrared Camera (LIR) detected a surprisingly large planetary-scale stationary gravity wave[21].

In this study, we mainly use data from the instrument IR2, the 2-μm infrared camera onboard Akatsuki, which images the planet between 1.74 to 2.32 μm[19,20,22] through four narrow band filters. IR2 started observation in March 2016. As the orientation of Akatsuki's elongated elliptical orbit is nearly fixed in the inertial space, times favored for day-side or night-side observations alternate in a Venus year. April, May, and June 2016 were the season for day-side observations, while March, July, and August, 2016 were favored for night-side observations. In this study, we use data obtained when the satellite was relatively close to Venus in March, July, and August, 2016. For comparison, we also use data from the instrument UVI, the ultraviolet imager of Akatsuki, which images the cloud top with scattered sunlight [19,20].

In this study, we use a novel automated cloud tracking method[23,24], which is more precise and more reliable than previously used methods[25]. The template size used is 7.5°×7.5°. To rule out the tracking of cloud features others than passive tracers (like atmospheric waves[21,24]), the results were manually verified by human eyes. Error evaluation is made with a novel method[23]. We also conduct independent manual tracking[26] to double-check. See Methods for more details.

Figure 1a shows the horizontal velocities (arrows) obtained from four 2.26 μm images taken at every two hours from 20:03 UTC, July 11, 2016. The orbiter at the time of observation was located near the longitude of sunset, and the wind is obtained between the local times 19:00 and 22:30 (see the abscissa of Fig. 2a and the tracer longitudes at end of cloud tracking shown in Fig. 1d). The wind field displayed in Fig. 1a is dominated by westward flow as reported in the previous studies with near-infrared observations[11--15]. However, the trajectories associated with the results ("+" marks) indicate that rotational speed peaks near the equator, which is also apparent in the clouds' displacements in high-pass filtered images (Fig. 1b-d).

The longitudinal average of zonal wind (Fig. 2b) displays a clear wind-speed maximum of around 85 m/s slightly to the north of the equator, which corresponds to a rotation period of about 5 days. It is not only the wind speed but also the angular speed around the rotational axis peaks near the equator, a feature previously unseen at any altitude. Accordingly, the angular momentum is also maximized there. Manual tracking independently confirmed this result. The present equatorial zonal flow is much faster than previously reported[11-15], but since it is found by tracking small-scale features, we can interpret that the jet feature is associated with the motion in the lower cloud layer.

Figure 2c shows the absolute vorticity computed from the longitudinal-mean zonal wind shown in Fig. 2b and the planetary vorticity; here, contribution of the meridional wind component, which is found small, is not included. Its absolute value is about $2\times10^{-5}$ s$^{-1}$ at 30°N and 20°S. If we naively compute its meridional gradient, it is roughly 1/4 of the "beta" value (meridional gradient of the vertical component of planetary vorticity) of the Earth.

Figure 3 shows zonal velocity obtained from dayside cloud-top images of reflected sunlight on July 11, 2016, corresponding to a region located about 80º eastwards (upstream of the mean zonal flow). Wind speeds from the UVI (365 nm, Fig. 3a) and IR2 (2.02 µm, Fig. 3b) are consistent with each other and exhibits a weak wind-speed minimum at around the equator, as sometimes observed[27]. The associated meridional gradient zonal absolute vorticity is roughly 1/10 of the Earth's beta value (not shown).

Figure 4 shows examples of mean zonal flow obtained with 2.26-µm IR2 nightside images at different occasions. During March 25, 2016 (Fig. 4a), zonal wind speeds are smaller than on July

11. Since the rotational angular speed does exhibit a peak, we do not call the low-latitudinal wind-speed maximum as the equatorial jet. The result is consistent with previous observations[11-15] within uncertainties. Nightside IR2 observations before July, 2016 are rather limited. A preliminary analysis for another day, April 15 (not shown), suggests that the zonal wind speeds near the equator are roughly between those on March 25 and July 11.

During August 2016, IR2 nightside observations are more frequent. The zonal wind at low latitude changes with time, but the equatorial jet seems persistent over August. A clear example can be found in Fig. 4b, which shows the results for August 13 and 15. Although there is a time difference of 2.5 days, corresponding to a half of the rotational period, equatorial jet features are quite similar, indicating their persistency. In contrast, Fig. 4c shows that zonal wind profiles obtained for August 25 and 26 have profiles markedly different compared to the other dates, and the latitude where wind speed is maximized are changed from the southern to the northern hemisphere during the interval of 1.5 days. Future wind measurements to complete the time coverage of August are expected to provide a more detailed insight about the transitions among the different wind profiles exhibited here.

The appearance and disappearance of the equatorial jet suggest that it may be subject to a type of long-term periodical event, what might also explain why it has not been reported in previous works. Actually, previous works are quite limited: Galileo/NIMS observed the Venus nightside only for some hours during the flyby[12], and the wind measurements for low latitude with Venus Express/VIRTIS are severely limited[13,14]. Ground-based observations helped to extend the time coverage, but measurement errors due to the low spatial resolution of the images precluded the identification of clear wind features[15]. Additional cloud-tracking using ground-based near-infrared observations with NASA's Infrared Telescope Facility (IRTF) in September, 2007 also suggest an equatorial westward flow faster than 80 m/s.

This study revealed that the zonal flow in the lower cloud layer has greater variability than was previously thought. The mechanism to create the equatorial jet is yet to be elucidated. The angular speed maximum indicates that it will not be created by a simple horizontal eddy diffusivity like the one supposed in theoretical studies[7,8,28], which smoothes angular velocity. Since the cloud-top level zonal wind is faster, downward momentum transport could create the equatorial jet. However, it is not straightforward to occur, since the upper cloud layer has stable stratification[29], and downwelling is not expected at low latitude. Vertical or horizontal momentum transport by atmospheric waves is a candidate. Interestingly, a numerical study[10] showed that atmospheric thermal tides created weak equatorial jets in the middle to upper cloud layer. Similar and more pronounced equatorial jets were found recently in a Venus general circulation model[30].

The mechanism of the superrotation in the Venusian atmosphere can be separated into two basic unknowns: one, the long-term torque balance between the whole atmosphere and the solid part of Venus, and, the other, the angular-momentum redistribution within the atmosphere to create a superrotation much faster than planetary rotation. The present discovery of the equatorial jet is expected to provide valuable hints and challenges on the latter aspect.

**Acknowledgements**

We deeply thank numerous (countless!) people who contributed to create and operate Akastuki. This study is supported by the following grants: JSPS KAKENHI 15K17767, 16H02225, and 16H02231, 16K17816; NASA Grant NNX16AC79G; JAXA's International Top Young Fellowship (ITYF).

**Author contributions**

T.H. developed automated cloud tracking and error evaluation methods, corrected the bore-sight of IR2 nightside images, conducted tracking, and interpreted the results. S.M. and K.O. contributed cloud tracking programing. T.S, T.M.S., K.S, T.I., and M.N. conducted IR2 observations and contributed to the operation of Akatsuki and observation planning. K.O, T.K., H.K., and M.T. developed the bore-sight correction applied to dayside images, and they also developed geographical mapping. J.P. conducted manual tracking with IR2 data based on independent geographical mapping. T.S, J.P., T.K., S.S.L., M.T. and E.F.Y. helped scientific interpretation and the review of previous studies. S.W., M.Y., and A.Y. conducted UVI observations. E.F.Y. conducted IRTF observations and the tracking with them.

**Supplementary animation GIF file**

Unfiltered two-hourly 2.26-μm radiance observed with IR2 from 12 h UTC, July 11 to 2 h UTC,

July 12, 2016. It is false-colored as shown at the bottom (W m$^{-2}$ sr$^{-1}$ μm$^{-1}$). The data are map-projected as if it were observed from an infinite distance at the longitude of 355° and the equator. The original images between 18 and 22 h include the dayside, but the white-out regions are not entirely saturated, since we set color-scales for good nightside representation. The other original images were obtained by removing the dayside from the field of view.

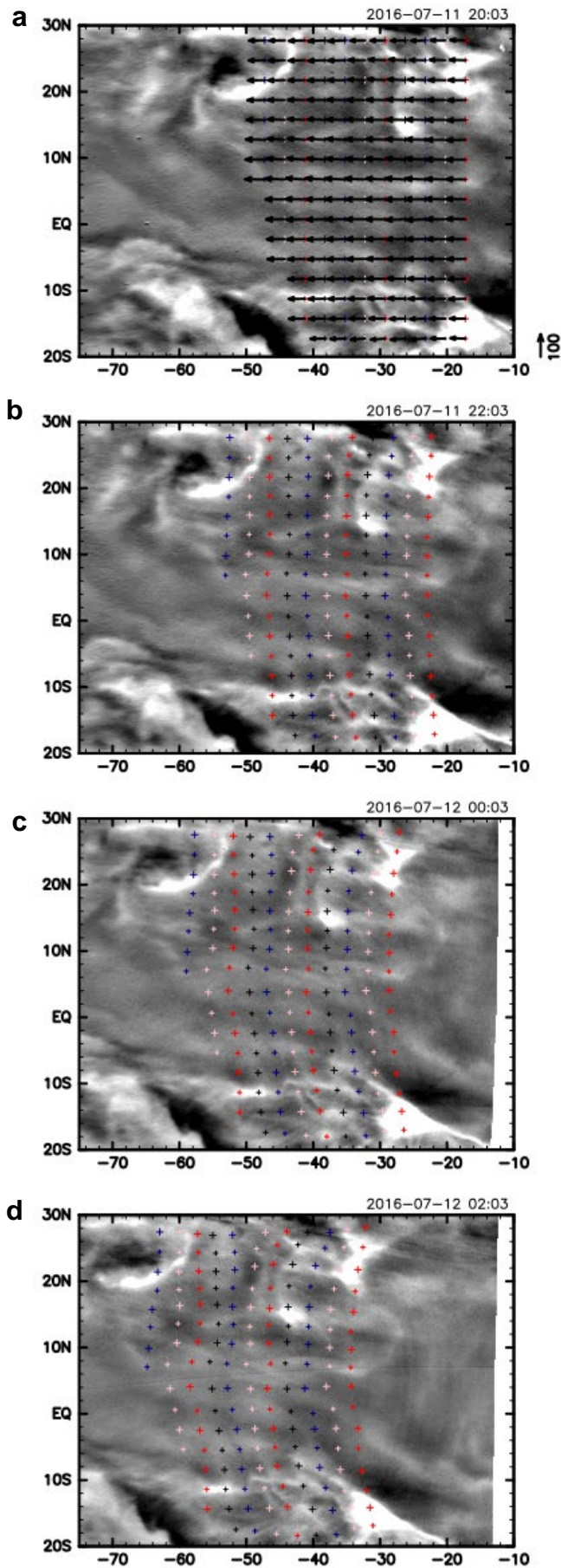

Figure 1: High-pass filtered radiance at 2.26 μm, horizontal velocities and associated trajectories for July 11-12, 2016. The abscissa is longitude, and the ordinate is latitude (limited to 20S-30N to focus on the equatorial jet). Grey-scale shading (a-d) represents the four two-hourly nightside IR2 images from 20:03 UTC, July 11 to 02:03 UTC, July 12, used for cloud tracking. The radiance is high-pass filtered by the two-dimensional Gaussian filter whose half-width at half-maximum is 4° in both longitude and latitude, which is similar to the filtering made in the pre-process of cloud tracking (see Methods). The grey-scale covers values from −0.02 (black) to 0.02 (white) W m$^{-2}$ sr$^{-1}$ μm$^{-1}$. Arrows (a) show the horizontal velocities whose scale in m/s is shown on the right-hand side. Colored "+" symbols at the initial time (a) show the centers of the template regions for cloud tracking (see Methods), while those at later times (b-d) show their positions advected by the velocities shown in a.

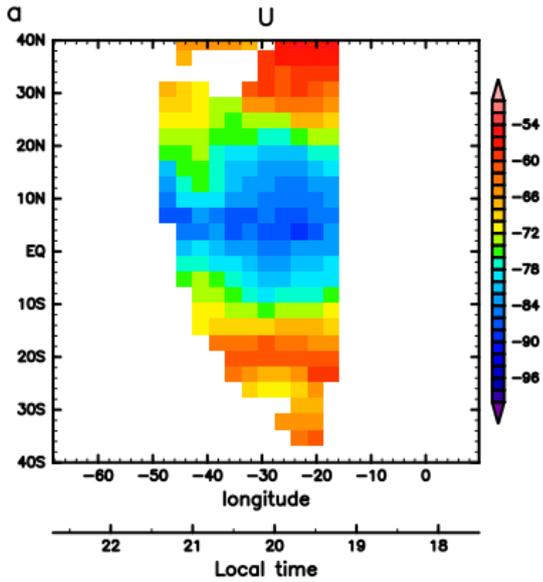
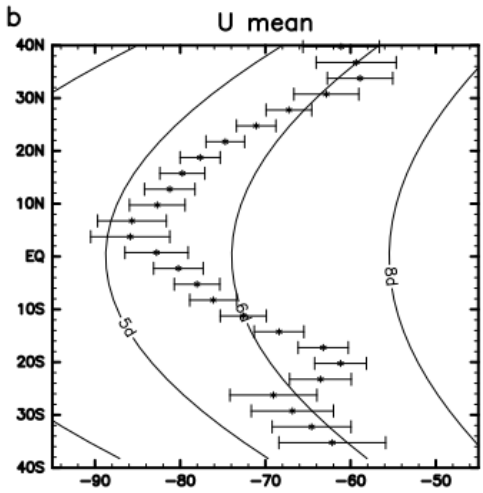
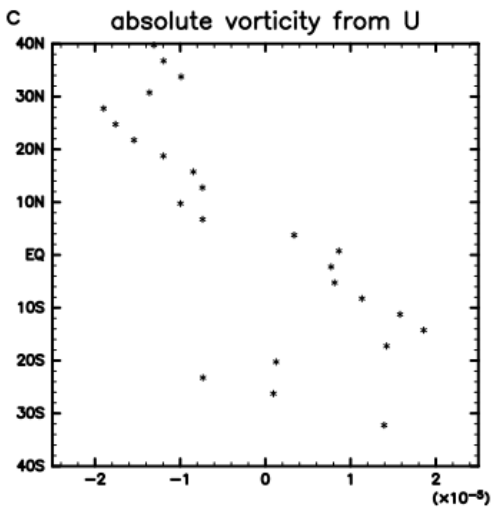

**Figure 2: Zonal wind and vorticity in the lower cloud layer in July 11-12, 2016.** Zonal (east-west) wind is defined as positive if it is eastward. **a**: Longitude (as well as local time)-latitude plot of zonal wind (m/s) in Fig. 1a. **b**: Longitudinal average of the zonal wind in **a**. The error bars show the uncertainty limit, derived from the sharpness of cross-correlation peaks and estimated pointing accuracy (see Methods). Curves show the corresponding rotation periods: 4, 5, 6, and 8 Earth days from left to right. **c**: absolute vorticity ($s^{-1}$) derived only from the mean zonal wind in **b** and the planetary vorticity.

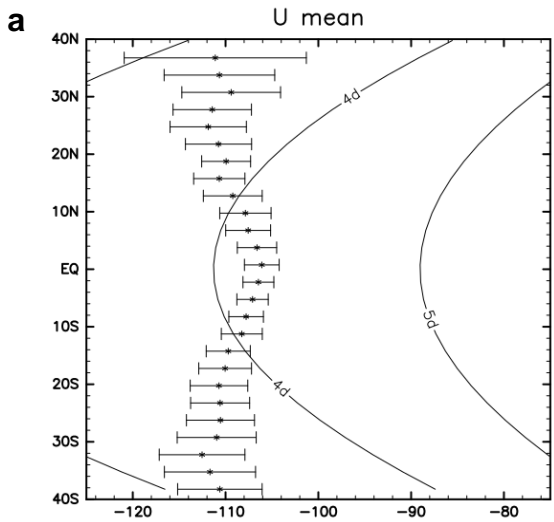

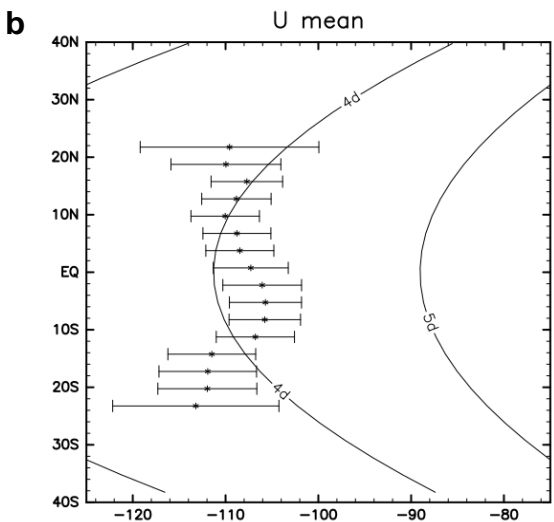

Figure 3: Zonal winds at the cloud top derived from UVI (365 nm) and IR2 (2.02 μm) data for July 11, 2016. a: Derived from three images obtained by UVI at 7, 9, and 11 h UTC, July, 2016 with the 365-nm filter; b: derived from the three images obtained by IR2 at 7, 9, and 11 UTC, July, 2016 with the 2.02-μm filter. Curves show the corresponding rotation periods: 3, 4, and 5 Earth days from left to right. See the caption of Fig.2b for error bars.

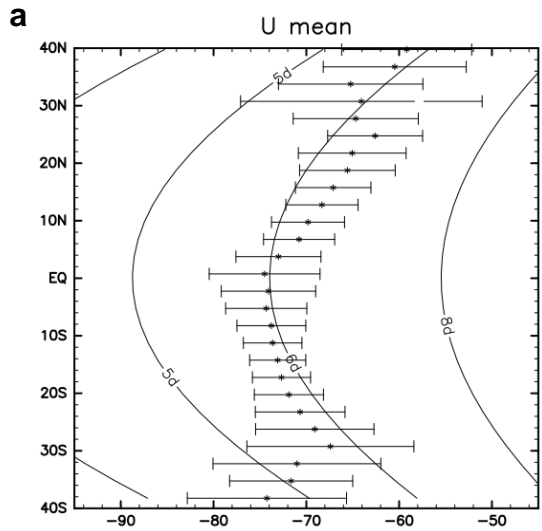

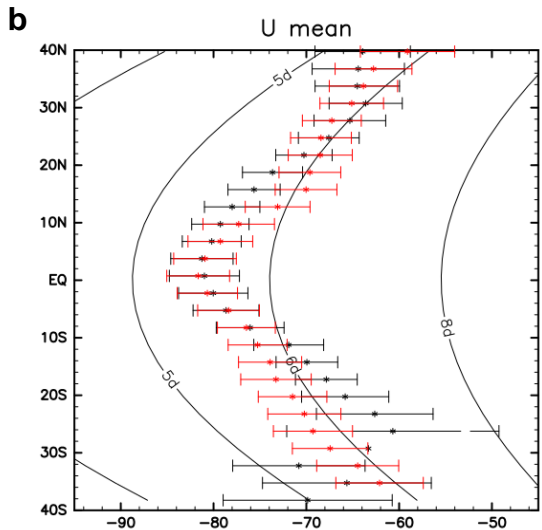

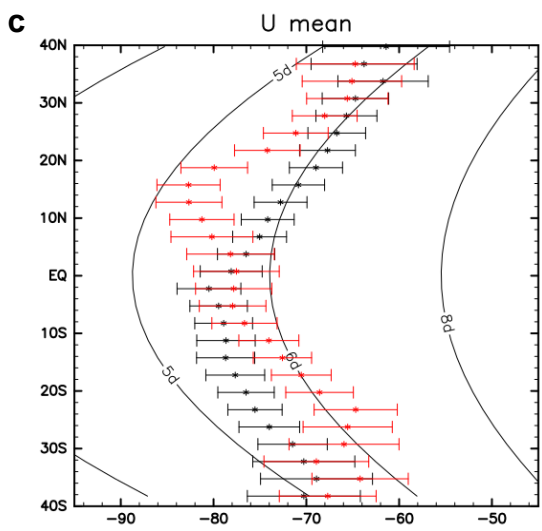

**Figure 4: Zonal winds in the lower cloud at various times.** The plot is as in Fig. 2b, but for the results of cloud tracking using images obtained on (a) March 25, (b) August 13 (black) and 15 (red), and (c) August 25 (black) and 26 (red). The local time ranges covered are approximately 0 to 4 h (March 25), 21:30 to 1:30 (August 13), 18:30 to 23 h (August 15), 18:30 to 23 h (August 25), and 19:30 to 24 h (August 26); the coverages for August 25 and 26 are in terms of the equator to southern hemisphere, and it is narrower in the northern hemisphere.

**Methods**

Automated cloud tracking

We used the version 0.2.1 IR2 nightside data, the version 0.2.1.1 IR2 dayside data, and the version 0.2.2.1 UVI data. The dayside image data are mapped onto a longitude-latitude grid with a resolution of 0.125° after correcting the bore-sight direction from the original navigation data by limb fitting[31], and a correction of the directional dependence of sunlight reflection is applied[31]. The nightside data are mapped onto the same grid, but the bore-sight direction from the original navigation data is corrected by maximizing the radial (inward, with respect to the sub-spacecraft point) component of the gradient of radiance. Before conducting cloud tracking, all the gridded data are applied with a two-dimensional band-pass filter with Gaussian functions with the sigma values 0.25° for low-pass and 3° for high-pass with latitude and 0.25°/cos$\varphi$ for low-pass and 3°/cos$\varphi$ for high-pass with longitude, where $\varphi$ is latitude[24] (the cos$\varphi$ factor is to compensate the length change in the circles of latitude, which is not applied in Fig. 1).

Cloud tracking was conducted with an enhanced cross-correlation method, where multiple cross-correlation surfaces are combined to estimate the horizontal velocity at a specific location[23], and a novel error correction/elimination method is applied[24]. Specific settings are as in our latest paper[24] except for the template size and the number of images used along with their time interval, which depend on quality considerations and data availability.

The size of the template images (sub-regions used to compute cross-correlation) is 7.5°×7.5° in this study; we checked the consistency among the results with smaller template sizes. Derived cloud-motion vectors are defined at the center of the template region at the initial time (see Fig. 1a); the wind derivation is conducted every 3° (Fig. 1a). Note that we use the spatial superposition of cross-correlation surfaces associated with nearby template regions[23,24]. As a result, a wind vector represents motion over an area greater than the template size, say 12°. No smoothing is applied to derived vectors.

The obtained velocity vectors are screened to retain only the ones with the peak correlation-coefficient values greater than 0.6 and the measures of precision, termed ε in our previous works[23,24] (computed from the sharpness of cross-correlation surfaces; see below for a further explanation) smaller than 15 m/s for nightside tracking (10 m/s for dayside tracking). Also, a post-process to eliminate inconsistent results is applied[24]. The results are verified by human-eyes in a way similar to conduct manual tracking.

The actual images used are as follows: four two-hourly images for IR2 nightside tracking starting at 20 h UTC, July 11, 0 h UTC, August 13, 11 h UTC, August 15, 4 h UTC, August 25, or 16 h UTC, August 26. The results presented in this study are based on the images obtained with the 2.26-μm filter, but we also conducted cloud tracking with the other two filters for nightside (1.74 and 2.32 μm) and confirmed that the results are consistent with each other. As for the nightside tracking for March 25, we only used two 2.26-µm images at 7 and 11 h UTC, since these two are the only available 2.26-µm images on the day. We also have another image at 7 h UTC with the 1.74-µm filter, so we conducted cloud tracking with this image and the 2.26-µm image at 11 h UTC for comparison. The two tracking results are remarkably consistent with each other. Dayside cloud tracking was conducted using three two-hourly images obtained at 7, 9, 11 h UTC, July 2016 for each of UVI 365-nm and IR2 2.02-µm images.

### Error estimation of the automated cloud tracking

We estimated the uncertainty of the mean zonal winds shown in Figs. 2a, 3, and 4 as follows:

$$e = (\langle \varepsilon_u^2 \rangle / N + b^2)^{1/2}.$$

The error bars in the figures show $\pm e$ of the estimated mean wind. Here, $\varepsilon_u$ represents the zonal-direction contribution to the parameter ε mentioned above (specifically, $\varepsilon_u$ is the greatest among $\varepsilon_{1u}$ and $\varepsilon_{2u}$ defined in our previous work[23], but it is very close to ε in most cases). These ε values are derived from the lower confidence bound of cross-correlation at the significance level of 90 %[23]. To evaluate the uncertainty of zonal mean, $\varepsilon_u$ at each wind-vector grid point is first binned over four grids (a bin consists of four consecutive grid points) by considering the oversampling as stated above. Their square values are averaged along longitude, which is expressed as $\langle \varepsilon_u^2 \rangle$. It is then divided by the number of bins, $N$, at each latitudinal grid, which represents the effective degrees of freedom. The term $b$ represents possible uncertainty due to incorrect positioning. We have confirmed that it is negligible for dayside images (manuscript in preparation), so it is simply set to zero. We estimated the uncertainty for the IR2 nightside images as described in the next paragraph. The resultant $b$ values are generally much smaller than the $[\langle\varepsilon_u^2\rangle/N]^{1/2}$ values, but it is not necessarily negligible.

The planetary extent in the IR2 nightside images is sometimes obscure, so careful treatment is needed to evaluate the accuracy of the pointing correction. The task was undertaken by comparing cloud tracking results conducted with two images at the time interval of two hours (for example, from two-hourly images between 8 and 16 h of a day, four wind estimates were obtained by using the image pairs taken at 8&10, 10&12, 12&14, and 14&16 h). We then computed the unbiased standard deviation of the areal mean winds equatorward of 30°N and 30°S within a day. The results over multiple days consistently suggested that the standard error associated with the combined six-hourly tracking conducted in this study is roughly equal to or smaller than 1 m/s (it is 3 m/s if two-hourly tracking is made). The error arises not only from the insufficient positioning correction but also from other factors. However, we simply attributed it to the former for safety. For consistency with the definition of ε, the standard error was expanded by a factor of 1.7 to comply with the 90-percentile. Therefore, we set $b = 1.7$ m/s for the six-houryly tracking (July and August, 2016) and $b = 2.5$ m/s for the four-hourly tracking (March 25, 2016).

**Manual cloud tracking**

A manual cloud tracking was conducted following an earlier study[26] for the nightside IR2 observation in July 11—12, 2016. The tracking was done independently, and we did not share the pre-processes including bore-sight correction.

**Cloud tracking with the 2007 IRTF data**

The cloud tracking was made with two methods; one is the automated tracking using the computer program described on a Website (http://www.skycoyote.com/FITSFlow/), and the other is manual tracking by a human operator.

31. Ogohara, K. et al. Automated cloud tracking system for the Akatsuki Venus Climate Orbiter data. *Icarus*, 217, 661-668 (2012).